\documentclass[sigconf]{acmart}

\usepackage{soul}
\usepackage{xcolor}
\usepackage{xspace}
\usepackage{url}
\usepackage{graphicx}
\usepackage{listings,multicol}
\usepackage[caption=false]{subfig}  
\usepackage[export]{adjustbox}
\usepackage{amsmath}

\usepackage{amssymb}
\usepackage{textcomp}
\usepackage{breakurl} 
\usepackage{listings}
\usepackage{hyperref}
\usepackage{cleveref}
\usepackage{algorithm}
\usepackage[noend]{algpseudocode}
\usepackage{dblfloatfix} 
\usepackage{enumitem}
\usepackage{kantlipsum}
\usepackage{multirow}
\usepackage{array}
\usepackage{booktabs}
\usepackage{tablefootnote}
\usepackage{color}
\usepackage{colortbl}

\usepackage{balance}
\usepackage{tikz}

\usepackage{tabularx}

\graphicspath{ {./images/} }

\makeatletter
\newcommand\footnoteref[1]{\protected@xdef\@thefnmark{\ref{#1}}\@footnotemark}
\makeatother

\newcommand{\companyName}{\textit{JetBrains}\xspace}

\newcommand{\totalRawWithoutDisqualifiedResponses}{18,032\xspace}

\newcommand{\totalFinalResponses}{14,396\xspace}

\newcommand{\totalQuestionsNum}{87\xspace}

\newcommand{\totalOpenEndedQuestionsNum}{8\xspace}

\newcommand{\totalQuestionsNumWithoutOpenEnded}{79\xspace}

\newcommand{\totalCategoriesNum}{10\xspace}

\newcommand{\totalCountriesNum}{173\xspace}

\newcommand{\rqOne}{What frustration factors do students
identify when studying CS?\xspace}

\newcommand{\rqTwo}{What formats are the most popular for learning computer science?\xspace}

\newcommand{\rqThree}{What are the learners’ attitudes towards in-IDE learning?\xspace}

\newcommand{\RQOneMostChallengingTotal}{14,780\xspace}

\newcommand{\RQOneQuitTotal}{10,666\xspace}

\newcommand{\RQOneOvercomingTotal}{8,182\xspace}

\newcommand{\RQOneHelpTotal}{14,799\xspace}

\newcommand{\RQThreeMostChallengingCloseTotal}{1,309\xspace}

\crefformat{footnote}{#2\footnotemark[#1]#3}

\newcommand{\zenodoLink}{ https://zenodo.org/records/16754164}
\title[Everything You Need to Know About CS Education: Open Results from a Survey of More Than 18,000 Participants]{Everything You Need to Know About CS Education: \\ Open Results from a Survey of More Than 18,000 Participants}

\author{Katsiaryna Dzialets}
\affiliation{
  \institution{JetBrains Research}
  \city{Munich}
  \country{Germany}
}
\email{katsiaryna.dzialets@jetbrains.com}

\author{Aleksandra Makeeva}
\affiliation{
  \institution{JetBrains}
  \city{Yerevan}
  \country{Armenia}
}
\email{alexandra.makeeva@jetbrains.com}

\author{Ilya Vlasov}
\affiliation{
  \institution{JetBrains Research}
  \city{Belgrade}
  \country{Serbia}
}
\email{ilya.vlasov@jetbrains.com}

\author{Anna Potriasaeva}
\affiliation{
  \institution{JetBrains Research}
  \city{Belgrade}
  \country{Serbia}
}
\email{anna.potriasaeva@jetbrains.com}

\author{Aleksei Rostovskii}
\affiliation{
  \institution{JetBrains Research}
  \city{Munich}
  \country{Germany}
}
\email{aleksei.rostovskii@jetbrains.com}

\author{Yaroslav Golubev}
\affiliation{
  \institution{JetBrains Research}
  \city{Belgrade}
  \country{Serbia}
}
\email{yaroslav.golubev@jetbrains.com}

\author{Anastasiia Birillo} 
\affiliation{
  \institution{JetBrains Research}
  \city{Belgrade}
  \country{Serbia}
}
\email{anastasia.birillo@jetbrains.com}

\renewcommand{\shortauthors}{Katsiaryna Dzialets et al.}


\AtBeginDocument{%
  }

\copyrightyear{2025}
\acmYear{2025}
\setcopyright{acmlicensed}\acmConference[CompEd 2025]{Proceedings of the ACM Global Computing Education Conference 2025 Vol 1}{October 21--25, 2025}{Gaborone, Botswana}
\acmBooktitle{Proceedings of the ACM Global Computing Education Conference 2025 Vol 1 (CompEd 2025), October 21--25, 2025, Gaborone, Botswana}
\acmDOI{10.1145/3736181.3747133}
\acmISBN{979-8-4007-1929-5/2025/10}

\begin{document}
    \begin{abstract}

Computer science education is a dynamic field with many aspects that influence the learner's path. While these aspects are usually studied in depth separately, it is also important to carry out broader large-scale studies that touch on many topics, because they allow us to put different results into each other's perspective. Past large-scale surveys have provided valuable insights, however, the emergence of new trends (\textit{e.g.}, AI), new learning formats (\textit{e.g.}, in-IDE learning), and the increasing learner diversity highlight the need for an updated comprehensive study. To address this, we conducted a survey with \totalRawWithoutDisqualifiedResponses learners from \totalCountriesNum countries, ensuring diverse representation and exploring a wide range of topics --- formal education, learning formats, AI usage, challenges, motivation, and more. This paper introduces the results of this survey as an open dataset, describes our methodology and the survey questions, and highlights, as a motivating example, three possible research directions within this data: challenges in learning, emerging formats, and insights into the in-IDE format. The dataset aims to support further research and foster advancements in computer education. 

\end{abstract}
    
    \maketitle
    
    \section{Introduction}
\label{section:introduction}

Computer science education is a large and complex field that provides many different and crucial areas for research --- from the reasons students pursue or quit programming~\cite{carter2006students, pappas2016investigating, petersen2016revisiting, kori2016factors} to different approaches and ways to learn the same topic~\cite{birillo2024bridging, hou2024effects, hew2014students}. Usually, these different areas are studied separately, allowing for a deep exploration of individual topics. However, it is also crucial to carry out broader, more general studies that touch on all the different sides of computer science education, because this allows us to put the results into perspective, to find higher-level trends and relationships in the data, which can inform further research. There are a lot of large-scale surveys that provide valuable insights like that~\cite{guzdial2012statewide, wang2016landscape, matzen2006teaching, hovey2019survey, budhiraja2024s}.
 
However, these studies suffer from certain threats to validity. Oftentimes, they are conducted with about 1,000 participants~\cite{hovey2019survey, matzen2006teaching} or are limited to one country, \textit{e.g.}, USA~\cite{wang2016landscape, guzdial2012statewide}, which does not provide a large enough picture of the field. Moreover, recent advancements like AI~\cite{valova2024students, liu2024teaching, prather2023robots}, new learning formats~\cite{birillo2024bridging}, and the growing diversity in learners~\cite{ibe2018reflections, dempster2020increasing} indicate a rapid change rate, 
entailing a strong demand for broader surveys that provide a high-level overview of the current state of computer science education. 

To fill this gap in research, we conducted a large-scale survey with two main focuses in mind: (1) a wide range of topics and questions, and (2) a diverse set of learners to ensure broad representation. 
Our survey contains a total of \totalQuestionsNum questions, organized into \totalCategoriesNum  categories, including: formal education, learning topics and formats, AI integration, learning challenges, motivation, and others. 
After carefully filtering and post-processing responses in accordance with industry standards~\cite{DevEcoJetBrains, developernatio, stackoverflowSurvey}, we ended up with responses of \totalRawWithoutDisqualifiedResponses participants from \totalCountriesNum countries.
This design and this scope allow us to explore complex patterns in the data and gain a bigger picture of modern computer science education.

Due to space constraints, the full analysis of the results is out of the scope of this paper, as it would require extensive detail and dozens of pages. Instead, we present the results as a publicly available dataset for the community. We also describe the used survey methodology and the detailed questionnaire.
Our goal is to enable researchers to utilize this dataset to test their hypotheses and conduct deeper studies that can advance the field of computer education.
\textbf{The dataset with all supplementary materials is available online: }\url{\zenodoLink}.

To showcase the usefulness of the dataset, we also outline three potential research directions: (1) challenges in learning, (2) emerging learning formats, and (3) practices in in-IDE learning.
To highlight interesting insights from these directions, we formulate three research questions:
\begin{enumerate}[leftmargin=1.3cm,start=1,label={\bfseries RQ\arabic*:}]
    \item \rqOne
    \item \rqTwo
    \item \rqThree
\end{enumerate}

These three example directions and research questions demonstrate the diversity and flexibility of our dataset: \textbf{RQ1} is more broad and pedagogical, \textbf{RQ2} has a bit narrower scope, and \textbf{RQ3} uses filtering to pinpoint a specific subset of respondents to target a smaller issue yet. We hope that this highlights how our dataset can be used by future researchers.
    \section{Background}
\label{section:motivation}

Since the goal of our survey is a broad understanding of computer science education, its background and related works are naturally very vast. For all the different aspects that we target in our study, detailed studies exist, for example, about the popularity of learning formats~\cite{pirker2014motivational, buerck2003learning, videnovik2024game}, the used AI tools~\cite{prather2023robots, valova2024students, budhiraja2024s}, the faced challenges~\cite{zhu2024k, yadav2016expanding}, etc. --- and we cannot hope to cover them all here. Because of this, we will describe several crucial works that were notable either for their scale or broadness.
 
Such works use different approaches to get insights, including surveys and interviews~\cite{hovey2019survey, wang2016landscape, guzdial2012statewide, kori2016factors, hou2024effects}, think aloud studies~\cite{oliveira2024investigating, pan2023insights, whalley2023think}, or the collection of the detailed data on interactions with programming environments or other tools~\cite{testMyCodeExploring, birillo2024one}. Survey studies tend to focus on answering general questions rather than delving deeper into the topic, because they can be carried out on a larger number of students than interviews or classroom studies. 

Wang et al.~\cite{wang2016landscape} conducted a large-scale survey involving over 15,000 students, parents, teachers, principals, and superintendents, uncovering disparities in access to and perceptions of K-12 computer science education in the United States. While the study provided valuable insights, its scope was limited to a single country. 

Several years later, a smaller-scale study was conducted with 821 computer science faculty members from 595 institutions across the United States~\cite{hovey2019survey}. This study explored the factors influencing CS faculty’s adoption of new teaching practices. The findings provide valuable insights for the CS community, such as the fact that student learning and engagement, along with practical constraints, play a significant role in shaping their decisions to implement new teaching methods. However, the number of participants is not very large, and the respondents are also limited to only one country.

Finally, with the increasing popularity of AI technologies, many studies have shifted focus to this area~\cite{valova2024students, hou2024effects, budhiraja2024s}. For instance, Valova et al.~\cite{valova2024students} recently conducted a comprehensive survey involving 102 high school and university students. This study examines the impact of ChatGPT on education by analyzing students' perceptions and usage patterns. In our survey, we also strive to ask some general questions about the AI usage and correlate them with other, more fundamental aspects of computer science education.

In summary, while previous large-scale surveys provide various insights, they often lack broadness of the topics and may be geographically limited. Aiming to inform future research in our quickly changing landscape of computer science education, we strive to carry out our survey with a broad set of questions and a diverse group of learners from as many different backgrounds as possible, helping researchers correlate their findings between different topics.  

    \section{Survey \& Dataset}
\label{section:methodology}

This section describes the methodology used to conduct our survey, namely: (1) survey design, (2) data collection, (3) processing open-ended responses, and (4) data weighting. At the end of the section, the resulting dataset is briefly summarized.

\subsection{Survey Design}

\textbf{Survey questions.} To provide a comprehensive overview of the Computer Science education field, we designed a large-scale survey with \totalQuestionsNum questions in total, organized into \totalCategoriesNum categories:

\begin{enumerate}
    \item \textbf{Demographics}. Basic demographic identifiers, including age, gender, geographical location, and language preferences.
    \item \textbf{Formal Education}. Educational background, institution types, and major of the study.
    \item \textbf{Career}. Career trajectory, industry experience, professional development, and job seeking behaviors.
    \item \textbf{Learning Topics and Formats}. Subject areas studied, self-assessed proficiency levels, educational platform preferences, and modality evaluations.
    \item \textbf{Coding Experience}. Technology adoption patterns and development practices.
    \item \textbf{Development Tools}. IDE usage patterns and tool preferences.
    \item \textbf{AI Integration}. AI tool usage in learning contexts.
    \item \textbf{Learning Challenges}. Obstacles encountered during learning processes and methods to overcome them.
    \item \textbf{Study Habits}. Learning routines, environments, productivity strategies, and device usage.
    \item \textbf{Motivation}. Drivers for Computer Science education engagement and career goals.
\end{enumerate}

The survey was designed by the authors together with industrial specialists in large-scale surveys from \companyName, a vendor of IDEs and tools for software development, who ensured the clarity of questions and answers, and piloted the survey internally with twenty people. The full overview of all the questions from the survey and the complete list of questions for each category can be found in the supplementary materials~\cite{supplementary}. 

\textbf{Data localization.} To ensure inclusivity and accommodate a diverse range of participants, the survey was available in ten languages: English, Chinese, French, German, Japanese, Korean, Brazilian Portuguese, Russian, Spanish, and Turkish.

\subsection{Data Collection}

\textbf{Data targeting.} Data collection was conducted from mid-February to the end of June of 2024. Participants were attracted via two separate approaches.  Firstly, potential respondents were reached through targeted advertisements on platforms such as X (formerly Twitter), Facebook, Bilibili, TikTok, and Instagram. Additionally, advertisements were placed on technology-focused community platforms, including Qiita, Quora, Reddit, Zhihu, and LinkedIn, encouraging participants to share the survey within their networks.

Secondly, we sent invitation emails to the list of people curated by \companyName who gave their consent for contact with research purposes.
During the data collection process we found that some regions are underrepresented, \textit{e.g.}, Japan, Ukraine, Russia, and Belarus. 
To address this, we employed external survey panels to get more responses from these regions. This practice is common in large-scale industry surveys~\cite{DevEcoJetBrains} to collect underrepresented responses. 

As a reward for participation, respondents could enter a raffle to win Amazon gift cards, subscriptions to educational platforms, and other prizes.

\textbf{Users privacy.} The \textit{General Research Terms and Conditions},\footnote{General Research Terms and Conditions: \url{https://www.jetbrains.com/legal/docs/terms/general-research-terms/}} which outlined the use of data for research purposes, were accepted by all participants. Prior to completing the survey, participants provided informed consent, and all personal data was processed in accordance with the privacy regulations. The terms explicitly stated that personal information would be anonymized following data collection, with personal identifiers retained only where necessary for validation purposes and then decoupled from analytical datasets during processing.

Based on the consent provided, we cannot share individual responses to open-ended questions, as these might contain potentially identifying information or perspectives that participants did not consent to share individually. However, as indicated in the agreement which participants signed, we can and do provide aggregated descriptions and thematic analyses of responses for each open-ended question, ensuring the privacy of individual respondents while still presenting valuable insights from the collected data (see details in Section~\ref{section:open-ended-responses-processing}).

\textbf{Filtering.} 
After closing the survey and receiving the initial responses, we applied several sanity and logical checks similar to other large-scale surveys from the industry~\cite{DevEcoJetBrains, stackoverflowSurvey}. These were aimed to remove \textit{suspicious} responses, \textit{e.g.}, if the participant indicated their age under 21 and at the same time more than 11 years of professional coding experience, or \textit{low-effort} responses, \textit{e.g.}, if the respondents answered the survey questions too quickly, with an average response time of less than five seconds per question, or selected more than nine different job roles. 
After this cleaning, we obtained the final \totalRawWithoutDisqualifiedResponses responses from \totalCountriesNum countries.
From them, \totalFinalResponses are \textit{full responses}, meaning that these respondents completed all questions that were asked of them in the survey.

\subsection{Processing Open-Ended Responses}
\label{section:open-ended-responses-processing}

Due to privacy reasons explained above, we processed the responses to open-ended questions to categorize them for inclusion into the final dataset. The processing pipeline consisted of two steps: (1)~translating responses into English and (2) coding the responses into categories.
For all the steps, we used the \texttt{GPT-4o} model, since it showed good performance on such tasks. 

\textbf{Translation.} First of all, we translated all the responses into one language --- English. This was done to unify the data and make further analysis easier.

\textbf{Categorization.} As the second step, we performed categorization to combine the responses with the same topic. 
For each question, we sent all of its responses to the LLM and asked it to define topics. Then, the first two authors manually checked the responses grouped into each topic and defined new ones based on the observations. In particular, the LLM sometimes grouped a lot of different things into ``Other'', and new topics were manually defined from there. When the categorization has been performed, the first two authors conducted the final manual check of the lists of topics.

After the categorization of each question, we simply replaced the open-ended responses with the resulting categories. For example, in question Q68 about frustrations (see Table~\ref{tab:rq-questions} further down in the paper), the open-ended responses were substituted with the corresponding categories, \textit{e.g.}, \textit{Taking breaks and physical activities}, \textit{Setting goals and reminding oneself of initial motivations}, \textit{Engaging in hobbies and personal projects}, and a few more. This methodology allows us to provide the general information from the open-ended questions without breaching the respondents' privacy.

\subsection{Data Weighting}
\label{section:data-weighting}

After the cleaning of the data, we followed industry standards~\cite{DevEcoJetBrains, developernatio, stackoverflowSurvey} to address potential sampling biases inherent in convenience sampling methods. 
During this step, we assigned a \textit{weight} to each response so that responses from underrepresented groups carry more influence during the analysis, while overrepresented groups are appropriately balanced. 
Incorporating these weights into the analysis is crucial to produce unbiased, representative results and improve the external validity of research. 
This weight was calculated in the following three steps:

\begin{enumerate}
    \item dividing responses into \textit{external} and \textit{internal};
    \item calculating weights for external responses;
    \item calculating weights for internal responses.
\end{enumerate}

\textbf{Dividing responses}.
Responses were defined into two categories by the target channel: \textit{external} --- those collected from social network ads, together with responses from peer referrals, and \textit{internal} --- responses collected from \companyName social networks and \companyName email list.

\textbf{Calculating weights for external responses}.
One crucial strength of our dataset is that the responses come from \totalCountriesNum countries, providing a rich and diverse perspective on computer science education. However, given how \textit{external} responses were collected, they might not correctly indicate the distribution of learners among different countries, and smaller and more remote countries with valuable perspectives may be underrepresented. Thus, the goal of weights for \textit{external} responses was to elevate responses from underrepresented countries. 

The correct distribution was taken from our previous internal research, which was aimed precisely at determining the distribution of software professionals in the world. It also showed that the distribution of learners is similar to that of professionals, which is why we can use this data. 
For each country, the weight was calculated according to this data, and then this weight was added to all \textit{external} responses from this country.

Summarizing, here are the main parameters of calculating weights for \textit{external} responses:

\begin{itemize}
    \item \textbf{Ground truth data}: a previous research of software professionals by \companyName, aimed at being representative around the world.
    \item \textbf{Equalized distribution}: the distribution of respondents across different countries.
    \item \textbf{Method}: post-stratification weighting~\cite{little1993post}.
\end{itemize}

\textbf{Calculating weights for internal responses}.
\textit{Internal} responses, in addition to suffering from the same bias in regards to countries, have additional biases --- the used tooling and the coding experience. While these respondents are not necessarily customers of \companyName, they still came from its resources and so might be biased to using the software developed by \companyName, and they are also usually more experienced in software engineering, as shown by other internal research. Thus, the goal of weights for \textit{internal} responses is to combat these three biases simultaneously.

For this, we can use the already scaled \textit{external} responses. Since they are not connected to \companyName in any way, we can assume that the distributions of the used tooling and experience there are not biased. At the same time, their weights calculated on the previous step take into account the representation of countries. The weight was calculated individually for each \textit{internal} response by solving a system of linear equations to balance the distributions as closely as possible: based on the country, the usage of tooling developed by \companyName, and coding experience.

Summarizing, here are the main parameters of calculating weights for \textit{internal} responses:

\begin{itemize}
    \item \textbf{Ground truth data}: \textit{external} responses from our survey, already scaled with regards to countries.
    \item \textbf{Equalized distributions}: the distribution of respondents by country, use of \companyName' tools, and coding experience.
    \item \textbf{Method}: QP-based calibration weighting~\cite{sarndal2007calibration}.
\end{itemize}

While the specific system of balancing cannot be shared due to the privacy policies of \companyName, all the calculated weights are added to the final dataset as a separate column and can be used for further analytics.

\textbf{Limitations}. Despite these measures, some bias may remain present, as the loyal audience of \companyName might have been more willing, on average, to complete the survey. Additionally, as the communities and learner ecosystem are constantly evolving, the possibility of some unexpected data fluctuations cannot be completely eliminated. We made heavy efforts to deal with these limitations and provide results that are extensive, rich, and balanced to industry standards.
    \subsection{Final Dataset}
\label{section:dataset}

As a result of the data collection and processing, we obtained the final \textbf{dataset with \totalRawWithoutDisqualifiedResponses respondents from \totalCountriesNum countries, including \totalFinalResponses fully completed responses}. For each participant, the dataset includes:

\begin{itemize}
    \item responses for \totalQuestionsNumWithoutOpenEnded single-choice, multi-choice, and ranking questions;
    \item categorized responses for \totalOpenEndedQuestionsNum open-ended questions (see Section~\ref{section:open-ended-responses-processing});
    \item the \textit{Weight} column to use in analysis (see Section~\ref{section:data-weighting});
    \item the \textit{Status} column to distinguish full responses from partial ones.
\end{itemize}

You can find the final dataset in the supplementary materials~\cite{supplementary}.

    \section{Research Directions}
\label{section:analysis}

In this section, we will highlight examples of what can be studied within our dataset. As we mentioned before, an exhaustive analysis is out of the scope of this paper, our goal is to show several interesting aspects, both general and specific, to motivate further research. In all the provided analyses, we considered both full and partial responses, and employed the weights described in Section~\ref{section:data-weighting}.

\subsection{Learning Challenges}

Many studies investigate learning challenges in computing education~\cite{pappas2016investigating, kori2016factors, hou2024effects, hew2014students}. Some of these works focus on the reasons for dropping out of computer science programs at a university~\cite{pappas2016investigating} or, conversely, why students decide to continue this learning path~\cite{kori2016factors}. Some papers examine the problems of less traditional modes of learning, for example, MOOCs~\cite{hew2014students} or in-IDE learning~\cite{birillo2025ide}. Finally, recent studies often focus on the challenges of using AI in programming courses and what kind of problems students and teachers might face with it~\cite{hou2024effects, valova2024students}. The data we have collected allows us to make analyses in this general direction and to understand what prevents students from learning. Our first research question is: \textbf{\textit{\rqOne}}

\textbf{Methodology.} To answer the first research question, we analyzed responses to four survey questions, written out in Table~\ref{tab:rq-questions}.
For Q67, we had \RQOneMostChallengingTotal multi-choice responses, for Q65 --- \RQOneQuitTotal multi-choice responses, for Q68 --- \RQOneOvercomingTotal open-ended responses, and, finally, for Q70 --- \RQOneHelpTotal multi-choice responses.

\textbf{Results}. Studying Computer Science presents several challenges, with the most common being \textit{understanding abstract and complex concepts} (51.39\%), followed by \textit{poor or missing documentation} (39.89\%), and \textit{getting stuck on problems} (39.39\%). The \textit{vastness of the field} (37.65\%) can feel overwhelming, while \textit{algorithmic problem-solving} (35.84\%) remains a tough skill to master. These challenges highlight the difficulty of both theoretical concepts and practical problem-solving in CS.

While studying CS is challenging, course design also plays a major role in student retention. The top reason for leaving a course is \textit{unengaging content} (51.19\%), followed by \textit{heavy workload and time constraints} (45.17\%). Many students struggle with courses that \textit{lack practical exercises} (30.01\%), while others find the material \textit{too simple} (25.54\%) or \textit{not relevant} (25.02\%). These factors suggest that beyond complexity, effective teaching methods and practical applications are crucial for keeping students engaged and motivated.

\begin{table}[t]
    \begin{tabularx}{\columnwidth}{ c  c X }
        \toprule

        \textbf{RQ} & \textbf{ID} & \textbf{Question} \\

        \midrule

        \multirow{4}{*}{\rotatebox[origin=r]{90}{Learning Challenges}}

        & \cellcolor{gray!30}67  & \cellcolor{gray!30}What do you find to be the most challenging thing about studying computer science? \\

        & 65  & Why did you quit the learning course or program? \\

        & \cellcolor{gray!30}68 & \cellcolor{gray!30}Do you have any methods of overcoming frustration and/or the feeling that you want to give up on your studies? \\

        & 70 & Where do you seek help when you don't understand something related to computer science? \\

        \midrule

        \multirow{3}{*}{\rotatebox[origin=r]{90}{Learning Formats}}

        & \cellcolor{gray!30}33  & \cellcolor{gray!30}Among the learning formats listed below, which ones do you have experience with when it comes to learning computer science topics? \\

        & 34 & Please rate your experience with these learning format(s) \\

        & \cellcolor{gray!30}35 & \cellcolor{gray!30}How familiar are you with the following MOOCs and code schools? \\

        \midrule

        \multirow{7}{*}{\rotatebox[origin=r]{90}{\makebox[115pt][c]{In-IDE Learning}}}

        & 16 & Over the past 12 months, have you studied computer science in any way? \\

        & \cellcolor{gray!30}33 & \cellcolor{gray!30}Among the learning formats listed below, which ones do you have experience with when it comes to learning computer science topics? \\

        & 35 & How familiar are you with the following MOOCs and code schools? \\

        & \cellcolor{gray!30}80 & \cellcolor{gray!30}What device do you mainly use for studying computer science? \\

        & 58 & Have you ever used an IDE for learning purposes? \\

        & \cellcolor{gray!30}49 & \cellcolor{gray!30}Which tool(s) do you use to run your code? \\

        & 55 & What do you use your IDE for? \\

        \bottomrule

    \end{tabularx}
    \vspace{0.2cm}
    \caption{Survey questions used for answering RQs.}
    \vspace{-0.5cm}
    \label{tab:rq-questions}
\end{table}

To overcome these frustration factors, students adopt various strategies. Here, we analyze the categories from an open-ended question, showcasing their usefulness. The most common approach is \textit{taking breaks and engaging in physical activities} (26.03\%), helping to reset focus and reduce stress. Others \textit{set goals and remind themselves of their initial motivations} (15.58\%) or \textit{practice self-reflection and adjust their mindset} (14.11\%) to stay on track. \textit{Seeking support from friends, family, or mentors} (7.03\%) and \textit{engaging in hobbies or personal projects} (6.93\%) also provide relief. However, 17.74\% have \textit{not yet found an effective strategy}, highlighting that many students continue to struggle with frustration and emphasizing the need for better support systems and coping techniques.

\begin{figure*}[t]
    \centering
    \includegraphics[width=0.95\textwidth]{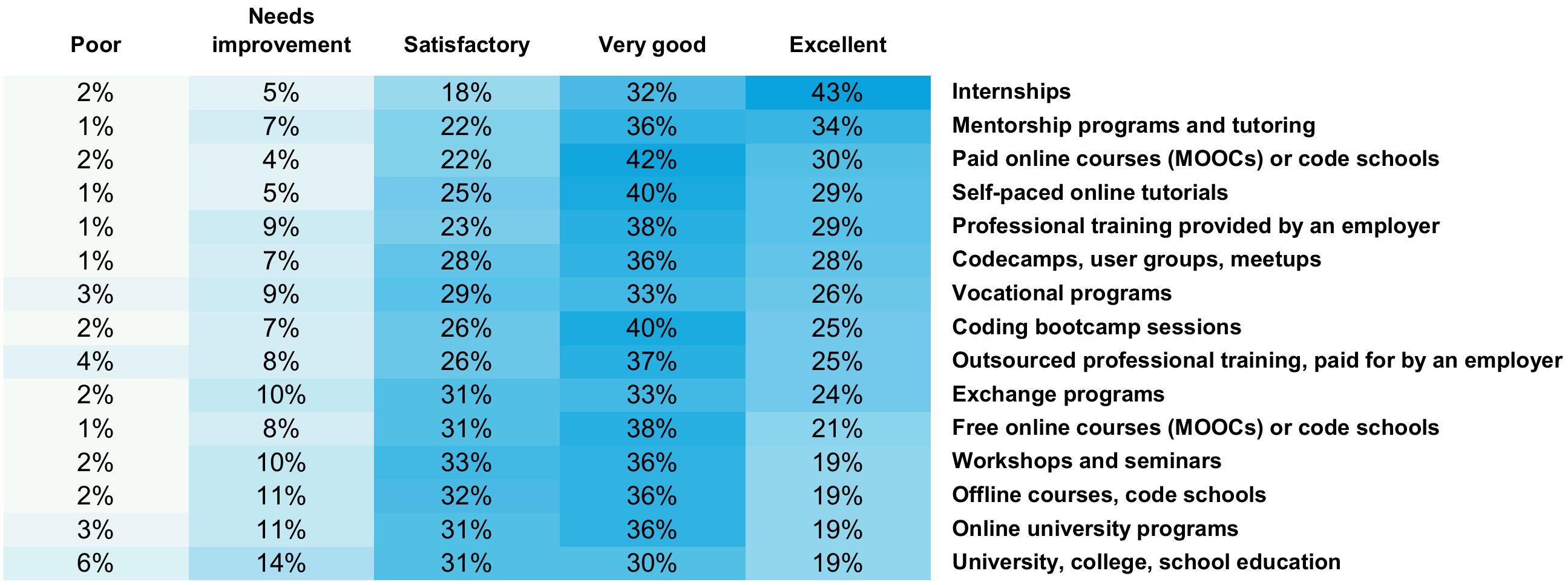}
 \vspace{-0.2cm}
    \caption{Answers to the question: \textit{``Please rate your experience with these learning format(s)''}.}
    \label{fig:formats}
    \vspace{-0.3cm}
\end{figure*}

When seeking help, \textit{Google} remains the most popular choice (74.79\%), but a significant number of respondents also use \textit{AI-based assistants} (61.24\%), highlighting the growing reliance on artificial intelligence for solving problems and clarifying concepts. \textit{Stack Overflow} (60.54\%) and \textit{YouTube} (51.68\%) follow closely, offering community-driven support and video tutorials. Additionally, \textit{friends and classmates} (42.84\%) are still valuable sources of help.

In summary, studying Computer Science involves challenges like abstract concepts, heavy workloads, and unengaging content, which can lead to frustration and course dropout. To cope, students often rely on digital tools, with AI-based assistants gaining popularity alongside traditional resources. This emphasizes the growing role of AI in providing real-time, personalized support for students navigating complex CS topics. It can be seen that our dataset can be used to answer the general questions about student behavior.

\subsection{Learning Formats}

Computer science education is a large field with many possible learning formats. Students can study programming in the traditional way at universities, with in-IDE courses~\cite{birillo2024bridging}, or with MOOCs~\cite{kruchinin2019investigation}.  
In this direction, somewhat more focused than the previous one, we answer the second research question: \textbf{\textit{\rqTwo}}

\textbf{Methodology.} To answer the second research question, we analyzed responses to three survey questions written out in Table~\ref{tab:rq-questions}.
Specifically, Q33 received 15,782 multiple-choice responses, Q34 received 15,737, and Q35 received 10,941.

\textbf{Results.} The survey results show that \textit{university, college, or school} remain the most popular learning format, chosen by 79.85\% of respondents. \textit{Self-paced online tutorials} follow closely at 66.08\%, highlighting a strong preference for flexible, independent learning. \textit{Free online courses and code schools} are also widely used (55.6\%), while \textit{internships} (29.92\%) and \textit{paid online courses} (28.28\%) are less common, implying cost and accessibility shape learning choices.

While \textit{university, college, or school} is the most popular learning format, Figure~\ref{fig:formats} shows that the respondents have the worst experience with it (19.85\% in total described their experience as ``Poor'' or ``Needs improvement''). Meanwhile, \textit{MOOCs and code schools} have a much better perception, offering flexibility and accessibility that challenge traditional education (only 5.97\% and 9.25\% in total described their experience as ``Poor'' or ``Needs improvement'' for paid and free options, respectively). This difference in perception suggests that students are turning to alternative educational platforms that better suit their needs. Finally, we can see that \textit{internships}, while not being very prevalent, provide the best experience, with 43.37\% of participants reporting them as ``Excellent''. 

Among the various MOOCs currently in use, \textit{Udemy} stands out as the most popular choice, with 29.08\% of respondents using the platform. \textit{Coursera} follows at 20.13\%, while \textit{JetBrains Academy} (16.15\%) and \textit{edX} (10.3\%) also attract a notable share of learners. \textit{Codecademy} (9.72\%) is slightly less common but still widely used. 

Overall, it can be seen that our dataset can be used to target a more narrow topic as well. In particular, it can be used to carry out meaningful comparisons between different questions, discovering discrepancies and overall trends. Similar to our analysis of the prevalence of different learning formats and respondents' experiences with them, we believe a lot of different topics in the dataset should be referenced with each other.

\subsection{In-IDE Learning}

In-IDE learning is a new and increasingly popular learning format~\cite{birillo2024bridging}. This format allows students to learn programming inside a professional IDE, not only to gain programming skills but also to become familiar with industry tools. Since the format is new and emerging, it is of interest to see which pain points it might currently have.
In this research direction, targeting a more narrow set of users that use a particular format, we answer the third research question: \textbf{\textit{\rqThree}}

\textbf{Methodology.}
Unlike in the first two RQs, in this one we need to first isolate a specific group of respondents. In-IDE learning courses are a subset of online courses, so this option is not explicitly present in the list of formats (see Figure~\ref{fig:formats}). Because of this, we need to select the participants who likely tried it using several questions. Of course, these results will be subject to threats to validity, but they are still interesting to inform further inquiries.

Based on the definition of \textit{in-IDE learning}~\cite{birillo2024one}, we focused on the questions specified in Table~\ref{tab:rq-questions}. We selected respondents who:
\begin{itemize}
    \item Q16: Recently studied CS \textit{(since the format emerged recently)};
    \item Q33: Had experience with MOOCs and online university programs;
    \item Q35: Currently use one of the following MOOCs: JavaRush or JetBrains Academy \textit{(MOOCs that support in-IDE learning)};
    \item Q80: Use laptops or desktops for studying;
    \item Q58: Use IDE for learning purposes;
    \item Q49: Use IDE for running code;
    \item Q55: Use IDE for purposes other than work and hobby.
\end{itemize}

In order to try to find out the challenges faced by respondents from this group, we examined \RQThreeMostChallengingCloseTotal responses to Q67. This is the same question we used for answering RQ1, which allows us to compare results between this group and all respondents.

\textbf{Results.} The top responses to this question closely align with those from RQ1, with \textit{understanding abstract and complex concepts} (55.9\%) remaining the most frequently reported challenge. \textit{Algorithmic problem-solving} (44.27\%) and \textit{poor or missing documentation} (43.57\%) also continue to be among the most commonly mentioned difficulties.
However, unlike in RQ1, the answer \textit{Hard to choose learning materials, courses, and platforms} (44.88\%) is now in the top. This might indicate that the field of in-IDE learning is not mature yet, and it is difficult for students to orient themselves in it and pick specific materials. This suggests a need for further research.

We can also use this research question to triangulate our data with previous studies. In our previous research~\cite{birillo2025ide} we conducted interviews with eight participants about their experience with in-IDE learning format, and one of the main findings was that the respondents find it difficult to set up a course environment. However, we cannot see this in our data: only 21.15\% of the filtered respondents reported \textit{operating system, compatibility, and environment issues}, and this percentage is very similar to that for all the respondents (22.79\%). 
Since our filtering is not exact, we cannot be sure that this result is final, but it can definitely be used to inform future research to study this issue more closely.

Overall, this research direction highlighted how our dataset can be used to carry out more specific analysis: use several questions to identify a particular group, see its answer to another question, and correlate that with another study. We believe that there are a lot of ways to apply this idea to our dataset in a variety of fields.

    \section{Conclusion \& Future Work}
\label{section:Conclusion}

In this paper, we share with the community the results of a broad large-scale survey that includes \totalRawWithoutDisqualifiedResponses respondents from \totalCountriesNum countries answering \totalQuestionsNum diverse questions about learning formats, challenges, motivation, using AI, and other topics. This data can be used to test hypotheses, answer different research questions, and allow researchers to discover trends in modern computer science education, thus potentially improving the teaching methods. We also share three potential research directions to exemplify how our dataset can be used for broad and narrow issues, by directly using its questions, correlating them with each other, or correlating them with previous studies. By sharing this dataset and explaining how it was created, we hope to give a starting point for new ideas and better practices in computer science education around the world. 

\section*{Acknowledgments}
We would like to thank the Strategic Research and Market Intelligence team at JetBrains --- especially Anastasia Molotova and Vladimir Volokhonsky from the Surveys subteam, --- as well as the JetBrains Academy team, with special thanks to Julia Amatuni and Maria Sharobaeva for making this survey possible.
    
    \bibliographystyle{ACM-Reference-Format}
    \balance
    \bibliography{ref}
\end{document}